\documentclass[conference]{IEEEtran}
\IEEEoverridecommandlockouts

\usepackage{mathrsfs}

\usepackage{amsmath}
\usepackage{ntheorem}

\usepackage{makeidx}  
\usepackage{algorithm}
\usepackage{algorithmic}
\usepackage{graphicx}
\usepackage{subfigure}
\usepackage{epstopdf}
\usepackage{bm}
\usepackage{cite}
\usepackage{stfloats}

\usepackage{amssymb}
\usepackage{graphicx}

\usepackage{url}

\usepackage{tabularx,booktabs}
\newcolumntype{C}{>{\centering\arraybackslash}X} 
\usepackage{lipsum}

\usepackage{makecell} 

\usepackage{graphicx}
\usepackage{color}
\newtheorem{thm}{Theorem}

\newtheorem{remark}{Remark}
\newtheorem{corollary}{Corollary}

\usepackage{diagbox}

\def\BibTeX{{\rm B\kern-.05em{\sc i\kern-.025em b}\kern-.08em
		T\kern-.1667em\lower.7ex\hbox{E}\kern-.125emX}}
	
\begin{document}

\title {Reconfigurable Intelligent Surface-Aided MISO Systems with Statistical CSI: Channel Estimation, Analysis and Optimization\\
\vspace{+5pt}
 \large\emph{ (Invited Paper)}}
\author{Kangda Zhi\textsuperscript{1}, Cunhua Pan\textsuperscript{1}, Hong Ren\textsuperscript{2}, Kezhi Wang\textsuperscript{3} and Maged Elkashlan\textsuperscript{1} \\
	
	\textsuperscript{1}EECS, Queen Mary University of London, \{k.zhi,  c.pan, maged.elkashlan\}@qmul.ac.uk \\
	\textsuperscript{2}National Mobile Communications Research Laboratory, Southeast University, hren@seu.edu.cn \\
	\textsuperscript{3}Department of Computer and Information Sciences, Northumbria University,  kezhi.wang@northumbria.ac.uk 
} 

\maketitle
\begin{abstract}
	This paper investigates the reconfigurable reflecting surface (RIS)-aided  multiple-input-single-output (MISO) systems with imperfect channel state information (CSI), where RIS-related channels are modeled by Rician fading. Considering the overhead and complexity in practical systems, we employ the low-complexity maximum ratio combining (MRC) beamforming at the base station (BS), and configure the phase shifts of the RIS based on long-term statistical CSI. 
Specifically, we first estimate the overall channel matrix based on the linear minimum mean square error (LMMSE) estimator, and evaluate the performance of MSE and normalized MSE (NMSE). Then, with the estimated channel, we derive the closed-form expressions of the ergodic rate. The derived expressions show that with Rician RIS-related channels, the rate can maintain at a non-zero value when the transmit power is scaled down proportionally to $1/M$ or $1/N^2$, where $M$ and $N$ are the number of antennas and reflecting elements, respectively. However, if all the RIS-related channels are fully Rayleigh, the transmit power of each user can only be scaled down proportionally to $1/\sqrt{M}$ or $1/N$.
Finally, numerical results verify the promising benefits from the RIS to traditional MISO systems.

\end{abstract}


\IEEEpeerreviewmaketitle

\section{Introduction}
As an emerging candidate for next-generation communication systems, reconfigurable intelligent surfaces (RISs), also called intelligent reflecting surfaces (IRSs), have attracted significant interest from both academia and industry\cite{pan2020reconfigurable}. By intelligently tuning the phase shifts of the impinging wave, an RIS can constructively strengthen the desired signal power or deconstructively weaken the interference signal, which yields an appealing passive beamforming gain\cite{wu2019intelligent,pan2020multicell,pan2020intelligent}.

The gains from RISs highly rely on the knowledge of the channel state information (CSI). However, the accurate acquirement of the CSI in RIS-aided systems is a challenging task, since RISs are typically passive without RF chains.   
To cope with this challenge, researchers have turned their attention to designing the phase shifts of RISs based on statistical CSI \cite{han2019large,zhi2020power,zhi2020directLinks}. Thanks to the slowly changing nature of the statistical CSI, we only need to configure the phase shifts of the RIS once over a large time-scale. Accordingly, this greatly reduce the computational complexity, and reduce the overhead of channel estimation and feedback.

However, existing statistical-CSI based contributions mainly characterized the systems performance under the assumption that perfect instantaneous overall CSI is available at the base station (BS). In practice, channel estimation error is inevitable.
Only a few contributions have taken imperfect CSI into consideration when configuring  the RIS  based on statistical CSI\cite{gao2021LMMSE,al2020intelligent,papazafeiropoulos2021intelligent}. Specifically, the impact of imperfect CSI on multi-RIS single-user systems was evaluated in \cite{gao2021LMMSE}, where the channel estimation is performed for each individual channel.
Different from \cite{gao2021LMMSE}, the authors in \cite{al2020intelligent,papazafeiropoulos2021intelligent} investigated the system performance when considering the cascaded channel estimation instead of individual channels. 
However, the contributions\cite{al2020intelligent,papazafeiropoulos2021intelligent} only considered a simple channel model where the RIS-BS link is fully line-of-sight (LoS) while RIS-users links are fully non-LoS (NLoS). Although this model is mathematically tractable, it cannot reveal the fundamental impact of the scatter richness of the propagation environment on RIS-aided systems. With a variety of richness of scatter, the quality of channel estimation and the system capacity performance in RIS-aided systems will exhibit wide differences.

Motivated by the above, we focus on the uplink transmission of an RIS-aided multiple-input-single-output (MISO) system with imperfect CSI, where all the RIS-related channels follow Rician distribution. We first propose a low-complexity linear minimum mean square error (LMMSE) estimator to estimate the channels as a whole in RIS-aided MISO systems. Based on the estimated channel, we theoretically investigate the performance of the mean square error (MSE) and normalized MSE (NMSE). Second, with low-complexity maximum ratio combining (MRC) receiver, rigorous expressions for the uplink ergodic data rate are derived in closed-form. Based on the expression, we design the phase shifts of the RIS and then characterize the asymptotic behaviors of the rate, the power scaling laws, and the impact of Rician factors. Finally, numerical results are presented to validate the superiority of RIS-aided MISO systems over traditional RIS-free MISO systems.

\section{System Model}\label{section_2}
\begin{figure}
	\setlength{\abovecaptionskip}{0pt}
	\setlength{\belowcaptionskip}{-20pt}
	\centering
	\includegraphics[width= 0.4\textwidth]{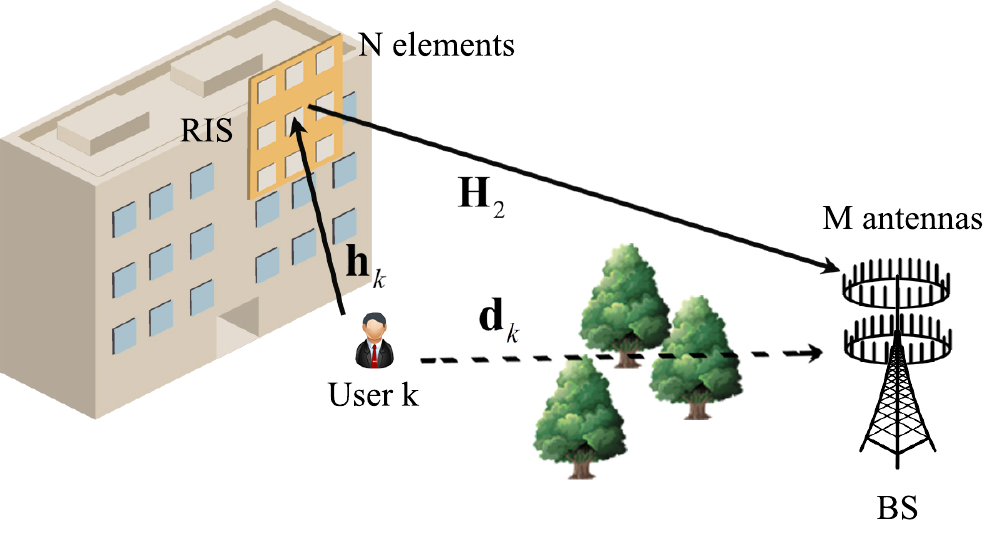}
	\DeclareGraphicsExtensions.
	\caption{An RIS-aided uplink MISO system.}
	\label{figure1}
	\vspace{-10pt}
\end{figure}
As illustrated in Fig. \ref{figure1}, we focus on the uplink transmission of an RIS-aided MISO system, where an RIS is deployed in the
proximity of a user to assist the transmission of a BS. The BS is equipped with $M$ antennas, the RIS is comprised of $N$ passive reflecting elements, and the user are equipped with single antenna. Without loss of generality, the user is denoted by user $k$. The channels from the user to the BS, from the user to the RIS, and from the RIS to the BS are denoted by $\mathbf {d}_k \in \mathbb{C}^{M\times 1}$,  $\mathbf {h}_k \in \mathbb{C}^{N\times 1}$ and $\mathbf {H}_2 \in \mathbb{C}^{M\times N}$, respectively.

The phase shift matrix of the RIS is expressed as $\mathbf{\Phi} = \mathrm{diag}\left\{e^{j\theta_1},...,e^{j\theta_N}\right\}$, where $\theta_n\in [0,2\pi)$ represents the phase shift of the $n$-th reflecting element. Based on the above definitions, the cascaded user-RIS-BS channel is obtained as $\mathbf{g}_k = \mathbf{H}_2 {\bf \Phi} {\bf h}_k$, and the overall channels is denoted by $\mathbf{q}_k = \mathbf{g}_k+\mathbf{d}_k \in \mathbb{C}^{M\times 1}$.
Then, the received signal vector at the BS is given by
$ \mathbf{y} = \sqrt{p} {\bf q}_k x_k +\mathbf{n} $ ,
where $p$ is the transmit power, $x_k$ represents the transmit symbol, and $\mathbf{n}\sim \mathcal{CN}\left({\bf 0},\sigma^2\mathbf{I}_M\right)$ denotes the vector of noise. 

Utilizing the channel estimate $\hat{\bf{q}}_k$, the BS performs MRC to process the received signal as
\begin{align}\label{rate_user_k}
{r}_k = \hat{\bf{q}}_k ^H \mathbf{y} = \sqrt{p} \hat{\bf{q}}_k  ^H \mathbf{q}_k {x}_k + \hat{\bf{q}}_k   ^H \mathbf{n}.
\end{align}

Since rich scatters often exist on the ground, as in \cite{han2019large}, we adopt the Rayleigh fading to model the direct channel as
$ \mathbf{d}_k = \sqrt{\gamma_k} \tilde{\mathbf{d}}_k $,
where $\gamma_k$ denotes the path-loss.
The entries of $\tilde{\mathbf{d}}_k$ are independent and identically distributed (i.i.d.) complex Gaussian random variables, and we have $ \tilde{\mathbf{d}}_k \sim \mathcal{CN}\left({\bf 0},\mathbf{I}_M\right)$.

Considering that the RIS is often installed at the facade of tall buildings and the BS also have a certain height, as in \cite{han2019large,zhi2020power,zhi2020directLinks}, we adopt Rician fading to model the RIS-related channels as
\begin{align}
\begin{array}{l}
	\mathbf{h}_{k}\!=\!\sqrt{\frac{\alpha_{k}}{\varepsilon_{k}+1}}\left(\sqrt{\varepsilon_{k}} \,\overline{\mathbf{h}}_{k}+\tilde{\mathbf{h}}_{k}\right), 
\mathbf{H}_{2}=\sqrt{\frac{\beta}{\delta+1}}\left(\sqrt{\delta} \, \overline{\mathbf{H}}_{2}+\tilde{\mathbf{H}}_{2}\right),
\end{array}
\end{align}
where $\alpha_k$ and $\beta$ are path-loss coefficients, $\varepsilon_k$ and $\delta$ are Rician factors. $\overline{\mathbf{h}}_k $ and $\overline{\mathbf{H}}_2 $ denote LoS components, whereas $\tilde{\mathbf{h}}_k $ and $\tilde{\mathbf{H}}_2 $ denote NLoS components. All the entries of NLoS components $\tilde{\mathbf{h}}_k $ and $\tilde{\mathbf{H}}_2 $ are i.i.d complex Gaussian random variables with zero mean and unit variance. For LoS components, we utilize the uniform squared planar array (USPA) model, and then $\overline{\mathbf{h}}_k $ and $\overline{\mathbf{H}}_2 $ are respectively expressed as
\begin{align}\label{uspa_hk}
&{\overline {\bf h}_k} =  {\bf a}_N \left( {\varphi _{kr}^a,\varphi _{kr}^e} \right),
& {\overline {\bf H}_2} = {\bf a}_M \left( {\phi _r^a,\phi _r^e} \right)    {\bf a}_N^H\left( {\varphi _t^a,\varphi _t^e} \right),
\end{align}
where $\varphi _{kr}^a$ ($\varphi _{kr}^e$) is the azimuth (elevation) angles of arrival (AoA) at the RIS from the user, $\varphi _t^a$  ($\varphi _t^e$) is the azimuth (elevation) angles of departure (AoD) from the RIS to the BS, and $\phi _r^a$ ($\phi _r^e$) is the azimuth (elevation) AoA at the BS from the RIS, respectively. Besides, $ {{\bf a}_X}\left( {\vartheta _{}^a,\vartheta _{}^e} \right) \in \mathbb{C}^{X\times 1}$ denotes the array response vector, whose $\rm x$-th entry is
\begin{align}\label{uspa}
\left[ {{\bf a}_X}\left( {\vartheta _{}^a,\vartheta _{}^e} \right) \right]_{\rm x} = 
&\exp\left\{j2\pi \frac{d}{\lambda }
	\left(    \lfloor \left({\rm x} - 1 \right)/\sqrt{X}\rfloor \sin \vartheta _{}^e\sin \vartheta _{}^a\right.\right. \nonumber \\
		& \left.\left.+ \left(\left(\mathrm{x}-1\right)\bmod \sqrt{X}\right)  \cos \vartheta _{}^e \right) \right\},
\end{align}
where $d$ and $\lambda$ denote element spacing and wavelength, respectively.
To simply the notations, in the sequel of this paper, we will use notations ${\bf \bar h}_k$, ${\bf a}_M$ and ${\bf a}_N$ to represent ${\bf a}_N \left( {\varphi _{kr}^a,\varphi _{kr}^e} \right)$, ${\bf a}_M \left( {\phi _r^a,\phi _r^e} \right)$ and ${\bf a}_N\left( {\varphi _t^a,\varphi _t^e} \right)$, respectively.

\section{Channel Estimation}\label{section_3}
Assume that the channel coherence interval consists of $\tau_c$ samples, and let $\tau$ denote the number of samples used  in the pilot transmission. 
 During the training part of each channel coherence interval, the user will send the pilot sequence to the BS. The transmit pilot power is $\tau p$, and the $1 \times \tau$ pilot sequence is ${\bf s}_k$, with ${\bf s}_k \mathbf{s}_k^H=1$. Therefore, the $M\times \tau$ pilot signal received by the BS can be written as
$ \mathbf{Y}_{p}=\sqrt{\tau p} \mathbf{q}_k\mathbf{s}_k+\mathbf{N} $,
where $\bf N$ denotes the $M\times\tau$ noise matrix whose entries are i.i.d. complex Gaussian random variables with zero mean and $\sigma^2$ variance. By multiplying $ \mathbf{Y}_{p} $ with $\frac{ {\bf s}^H_k} {\sqrt{\tau p}}$, the BS can obtain the observation vector for the user as
\begin{align}\label{y_p}
\mathbf{y}_{p}^{k}=\frac{1}{\sqrt{\tau p}} \mathbf{Y}_{p} \mathbf{s}_{k}^{H}=\mathbf{q}_{k}+\frac{1}{\sqrt{\tau p}} \mathbf{N s}_{k}^{H}.
\end{align}

To obtain closed-form channel estimates and enable our analysis, we will adopt the sub-optimal LMMSE estimator to perform the channel estimation.\footnote{All proofs in this paper are omitted due to the page limitation. Please refer to  \cite{zhi_mimo_imperfect} for details.}
\begin{thm}\label{theorem1}
	Using the observation vector ${\mathbf{y}}^k_{p}$, the LMMSE estimation $\hat{\mathbf{q}}_{k}$ for channel ${\mathbf{q}}_{k}$ is given by
$  \hat{\mathbf{q}}_{k}=\mathbf{A}_{k} \mathbf{y}_{p}^{k}+\mathbf{B}_{k} $,
	where
	\begin{align}
&\mathbf{A}_{k}=a_{k 3} \mathbf{a}_{M} \mathbf{a}_{M}^{H}+a_{k 4} \mathbf{I}_{M},\\
&\mathbf{B}_{k}=\left(\mathbf{I}_{M}-\mathbf{A}_{k}\right) \sqrt{c_{k} \delta \varepsilon_{k}} \, \overline{\mathbf{H}}_{2} \mathbf{\Phi} \overline{\mathbf{h}}_{k},\\
	&c_{k}= \frac{\beta \alpha_{k}}{(\delta+1)\left(\varepsilon_{k}+1\right)},\\
&a_{k1}=N c_{k} \delta,
\quad a_{k2} = N c_{k}\left(\varepsilon_{k}+1\right)+\gamma_{k}, \\
&a_{k 3}= \frac{a_{k 1} \frac{\sigma^{2}}{\tau p}}{\left(a_{k 2}+\frac{\sigma^{2}}{\tau p}\right)\left(a_{k 2}+\frac{\sigma^{2}}{\tau p}+M a_{k 1}\right)}, 
\; a_{k 4}= \frac{a_{k 2}}{a_{k 2}+\frac{\sigma^{2}}{\tau p}},
	\end{align}
	and the NMSE in the estimation of $\mathbf{q} _{k}$ is   
	\begin{align}\label{NMSE}
&\operatorname{NMSE}_{k}=\frac{\frac{\sigma^{2}}{\tau p}\left(M a_{k 1} a_{k 2}+a_{k 2}^{2}+\left(a_{k 1}+a_{k 2}\right) \frac{\sigma^{2}}{\tau p}\right)}{\left(a_{k 2}+\frac{\sigma^{2}}{\tau p}\right)\left(a_{k 2}+\frac{\sigma^{2}}{\tau p}+M a_{k 1}\right)\left(a_{k 1}+a_{k 2}\right)}.
	\end{align}
	
\end{thm}

As shown in Theorem \ref{theorem1}, we only estimate the overall channel ${\bf q}_k \in\mathbb{C}^{M\times 1}$, which possesses the same dimension as in traditional MISO systems. Therefore, we only require that the length of pilot sequences is larger than the number of users, i.e., $\tau\geq 1$. Compared with the estimation method which needs the estimation of all $MN$ individual channels\cite{9130088}, our method has very low overhead and complexity.

\begin{remark}\label{remark1}
We can see that the NMSE expression (\ref{NMSE}) is unrelated to the phase shifts of the RIS. 
 Therefore, the proposed LMMSE estimation method can work compatibly with the statistical CSI-based RIS design, since the phase shifts of RIS don't need to be adjusted in each channel coherence interval.
\end{remark}

\begin{corollary}\label{corollary1}
	The asymptotic NMSE performance is
	\begin{align}\label{NMSE_scale1}
\lim \nolimits_{  \frac{\sigma^{2}}{\tau p} \rightarrow 0} \mathrm{NMSE}_{k} \rightarrow 0, 
\quad \lim \nolimits_{N \rightarrow \infty} \mathrm{NMSE}_{k} \rightarrow 0.
	\end{align}
\end{corollary}

Note that the NMSE between $ 0 $ (perfect estimation) and $ 1  $ (using the mean value of the variable as the estimate) measures the relative estimation error per antenna \cite{bjornson2017massive}. In traditional MISO systems, a common method to reduce the NMSE is to increase the length of pilots, i.e., increasing $\tau$. However, Corollary \ref{corollary1} indicates that in RIS-aided MISO systems, increasing the number of reflecting elements of RIS can play a similar role as increasing $\tau$. Therefore, increasing RIS reflecting elements not only helps improve the system capacity, but also helps decrease the NMSE.

To better understand the impact of increasing $N$ on channel estimation, we present the following asymptotic results.
\begin{corollary}\label{corollary2}
When $\tau\to\infty$, we have $\hat{\mathbf{q}}_{k} \to {\mathbf{q}}_{k}$. By contrast, when $N\to\infty$, we have 
\begin{align}\label{LS_estimator}
&\hat{\mathbf{q}}_{k}\to {\mathbf{q}}_{k} + \frac{1}{\sqrt{\tau p}}  \mathbf{N s}_{k}^{H},
\quad {\rm MSE}_k \to \frac{\sigma^2}{\tau p} {\bf I}_M.
\end{align}

\end{corollary}

Although the NMSE will converge to zero with infinite $N$, Corollary \ref{corollary2} shows that different from increasing $\tau$, the MSE under LMMSE estimator will converge to a non-zero constant.  It is interesting to find that the asymptotic MSE in (\ref{LS_estimator}) is the same as the MSE performance of least-squares (LS) estimator\cite{bjornson2017massive}. It is well-known that the LS estimator has a worse estimation quality than the LMMSE estimator. Therefore, we can roughly deduce that the MSE performance under LMMSE is an increasing function of $N$ which converges to an upper bound of the LS estimator. This result will be validated in Section \ref{section_6}.


\section{Analysis of Uplink Ergodic Rate}\label{section_4}
As in \cite{bjornson2017massive}, we resort to the so called use-and-then-forget
bound to characterize the ergodic rate performance. We rewrite the received signal $r_k$ in (\ref{rate_user_k}) as
\begin{align}
r_{k}=\underbrace{\sqrt{p} \, \mathbb{E}\left\{\hat{\mathbf{q}}_{k}^{H} \mathbf{q}_{k}\right\} x_{k}}_{\text {Desired signal }}+\underbrace{\sqrt{p}\left(\hat{\mathbf{q}}_{k}^{H} \mathbf{q}_{k}-\mathbb{E}\left\{\hat{\mathbf{q}}_{k}^{H} \mathbf{q}_{k}\right\}\right) x_{k}}_{\text {Signal leakage }}
+\underbrace{\hat{\mathbf{q}}_{k}^{H} \mathbf{n}}_{\text {Noise }}.
\end{align}

Then, we present the closed-form expression of ergodic rate in the following theorem.

\begin{thm}\label{theorem2}
The ergodic rate of the user is lower bounded by
\begin{align}\label{rate}
\underline{R}_{k}=\frac{\tau_c-\tau}{\tau_c}\log _{2}\left\{1+\frac{  p E_{k}^{(\rm s i g n a l)}    }{p E_{k}^{(\rm l e a k a g e)}  +\sigma^{2} E_{k}^{(\rm {noise})}    }\right\},
\end{align}
with $ E_{k}^{(\rm signal)}  =  \left(E_{k}^{(\rm {noise})}\right)^{2}$,
\begin{align}\label{signal_LMMSE}
E_{k}^{(\rm {noise})}=M\times& \left\{    \left|f_{k}(\mathbf{\Phi})\right|^{2} c_{k} \delta \varepsilon_{k}+N c_{k} \delta e_{k 2}\right. \nonumber\quad\qquad\qquad\\
&+\left(N c_{k}\left(\varepsilon_{k}+1\right)+\gamma_{k}\right) e_{k 1}  \Big\},
\end{align}
\begin{align}
\begin{array}{l}
E_{k}^{(\rm leakage)} \\
=M\left|f_{k}(\mathbf{\Phi})\right|^{2} c_{k}^{2} \delta \varepsilon_{k}\left\{N\left(M \delta+\varepsilon_{k}+1\right)\left(e_{k 2}^{2}+1\right)\right.\\
\qquad\qquad\qquad\qquad\quad\left.+2\left(M e_{k 1}+e_{k 2}\right)\left(e_{k 2}+1\right)\right\} \\
+M\left|f_{k}(\mathbf{\Phi})\right|^{2} c_{k} \delta \varepsilon_{k}\left\{  \gamma_{k}+\left(\gamma_{k}+\frac{\sigma^{2}}{\tau p}\right) e_{k 2}^{2} \right\} \\
+M^{2} N^{2} c_{k}^{2} \delta^{2} e_{k 2}^{2} \\
+M N^{2} c_{k}^{2}\left\{2 \delta\left(\varepsilon_{k}+1\right) e_{k 2}^{2}+\left(\varepsilon_{k}+1\right)^{2} e_{k 3}\right\} \\
+M^{2} N c_{k}^{2}\left\{\left(2 \varepsilon_{k}+1\right) e_{k 1}^{2}+2 \delta e_{k 1} e_{k 2}\right\} \\
+M N c_{k}\Big\{    c_{k}\left(2 \delta e_{k 2}^{2}+\left(2 \varepsilon_{k}+1\right) e_{k 3}\right)  \\
\qquad\qquad\left.+\left(2 \gamma_{k}+\frac{\sigma^{2}}{\tau p}\right)\left(\delta e_{k 2}^{2}+\left(\varepsilon_{k}+1\right) e_{k 3}\right)\right\} \\
+M \gamma_{k}\left(\gamma_{k}+\frac{\sigma^{2}}{\tau p}\right) e_{k 3},
\end{array}
\end{align}

where
\begin{align}\label{f_k_Phi_definition}
&f_{k}({\bf\Phi}) = \mathbf{a}_{N}^{H} {\bf\Phi} \overline{\mathbf{h}}_{k}, \quad
e_{k 1} = a_{k 3}+a_{k 4} \in \left[0,1\right], \\
&e_{k 2} = M a_{3}+a_{4}  \in \left[0,1\right], \\
&e_{k 3} = M a_{k 3}^{2}+2 a_{k 3} a_{k 4}+a_{k 4}^{2} \in \left[0,1\right].
\end{align}

\end{thm}

The closed-form expressions in Theorem \ref{theorem2} do not rely on fast changing NLoS channel components, and do not involve computationally burdened integral and matrix inverse operators. Therefore, in contrast to time-consuming Monte-Carlo simulations, the evaluation of rate performance based on Theorem \ref{theorem2} can be computed rapidly. Besides, comparing the expressions in Theorem \ref{theorem2} with those in \cite[Lemma 1]{zhi2020power}, one can see that the impact of imperfect CSI is fully characterized by the parameters $e_{k1}, e_{k2}, e_{k3}$ and $\frac{\sigma^{2}}{\tau p}$. For the perfect CSI scenario without channel estimation error, we can assume $\tau=\infty$, which leads to $e_{k1}=e_{k2}=e_{k3}=1$ and $\frac{\sigma^{2}}{\tau p}=0$.

Next, we will design the phase shifts of the RIS. If $N=1$, the phase shifts can be set arbitrarily\cite{zhi_mimo_imperfect}. Therefore, we only consider the case $N>1$ in the following.
From Theorem \ref{theorem2}, we can see that the only phase shifts related variable is $\left|f_k\left(\mathbf{\Phi}\right)\right|^2$. We can treat $\left|f_k\left(\mathbf{\Phi}\right)\right|^2$ as a new variable $x$ where $0\leq x \leq N^2$\cite{zhi_mimo_imperfect}. Then, we can rewrite the numerator and denominator of SNR in Theorem \ref{theorem2} as $E_{k}^{(\text {signal})}=\left(s_{1} x+s_{2}\right)^{2}$ and $E_{k}^{(\text {leakage})}+\frac{\sigma^{2}}{p} E_{k}^{(\text {noise})}=t_{1} x+t_{2}$, respectively. The detailed expressions of $s_1, s_2, t_1, t_2$ are omitted here, since it can be readily obtained from Theorem \ref{theorem2}. As a result, the SNR in Theorem \ref{theorem2}  can be rewritten as
\begin{align}
\mathrm{SNR}_{k}\left(x\right)=\frac{\left(s_{1} x+s_{2}\right)^{2}}{t_{1} x+t_{2}}.
\end{align}

The rate can be maximized by maximizing the SNR. Thus,
we calculate the first-order derivative of $\mathrm{S N R}_k\left(x\right)$ with respect to $x$ as
\begin{align}
\frac{\partial \mathrm{SNR}_{k}\left(x\right)}{\partial x}=\left(s_{1} x+s_{2}\right) \frac{s_{1} t_{1} x+2 s_{1} t_{2}-s_{2} t_{1}}{\left(t_{1} x+t_{2}\right)^{2}}.
\end{align}

Note that $s_1,s_2,t_1,t_2>0$. Set the first-order derivative to zero. Under the constraint $x\ge 0$, the only possible solution is
\begin{align}\label{define_x0}
\frac{s_{2} t_{1}-2 s_{1} t_{2}}{s_{1} t_{1}} \triangleq x_{0}.
\end{align}

Therefore, the optimal phase shifts design can be decided by discussing the value of $x_0$, which is presented as follows.
\begin{thm}\label{theorem3}
	The phase shifts of RIS can be designed under the following four cases:
	\begin{itemize}
		\item Case 1: If $x_0\leq 0$, it is optimal to set $\left|f_k\left(\mathbf{\Phi}\right)\right|=N$. 
		
		\item Case 2:  If $x_0\ge N^2$, it is optimal to set $\left|f_k\left(\mathbf{\Phi}\right)\right|=0$.  
		
		\item Case 3:  If $0<x_0< N^2$, we should compare ${\rm SNR}_k\left(0\right)$ with ${\rm SNR}_k\left(N^2\right)$, and choose the maximum one.
		
		\item Case 4: If $N\to \infty$, it is optimal to set $\left|f_k\left(\mathbf{\Phi}\right)\right|=N$. 
	\end{itemize}
\end{thm}

Theorem \ref{theorem3} indicates that it is feasible to design the phase shifts as $\left|f_k\left(\mathbf{\Phi}\right)\right|=N$ if $N$ is large. Therefore, in the rest of this section, we will set $\left|f_k\left(\mathbf{\Phi}\right)\right|=N$ to facilitate our analysis.

Substituting Theorem \ref{theorem2} with $\left|f_{k}(\mathbf{\Phi})\right|=N$, we can see that the power of desired signal is in the order of $\mathcal{O}\left(M^2N^4\right)$, the power of signal leakage is in the order of $\mathcal{O}\left(M^2N^3\right)$, and the power of noise term is in the order of $\mathcal{O}\left(MN^2\right)$. Therefore, the rate is bounded when $M\to\infty$, but can grow unbounded as $N\to\infty$. Besides, we can reduce the power proportionally to $1/M$ or $1/N^2$ while keeping the rate non-zero as $M,N\to\infty$. 
\begin{corollary}\label{corollary_single_user_MNN}
If the transmit power is reduced to $p={E_u}/{N^2}$, as $N\to\infty$, the rate is lower bounded by
\begin{align}\label{rate_singleUser_scale_NN}
\underline{R}_{k}^{} \rightarrow \frac{\tau_{c}-\tau}{\tau_{c}} \log _{2}\left(1+\frac{E_{u}}{\sigma^{2}} M \frac{\beta \alpha_{k} \delta \varepsilon_{k}}{(\delta+1)\left(\varepsilon_{k}+1\right)}\right).
\end{align}
\end{corollary}

The SNR in (\ref{rate_singleUser_scale_NN}) is equal to that in $ M $-antennas systems with path-loss $c_k\delta\varepsilon_{k}$. Besides, the  SNR is uncorrelated with $\tau$ and the same SNR can be realized by applying the corresponding power scaling in perfect CSI-based systems\cite[(17)]{han2019large}. 
In other words, the increase of $N$ could compensate the detrimental effects of limited pilot lengths, and $\tau=1$ is adequate in  (\ref{rate_singleUser_scale_NN}).
Furthermore, it is easy to find that the rate is an increasing function of Rician factors $\delta$ and $\varepsilon_{k}$. Therefore, the LoS environment is beneficial for RIS-aided single-user systems.

However, we can find that the power scaling laws in (\ref{rate_singleUser_scale_NN}) no longer hold under fully Rayleigh case, i.e., $\delta=\varepsilon_{k}=0$. In this case, we can reduce the power proportionally to $1/\sqrt{M}$ or $1/N$ while keeping the rate non-zero when $M,N\to\infty$.
\begin{corollary}\label{corollary_single_user_rayleigh}
	Let $\delta=\varepsilon_{k}=0$. if the transmit power is reduced as $p={E_u}/{N}$, as $N\to\infty$, the rate is lower bounded by
	\begin{align}\label{scaling_N_rayleigh}
 \underline{R}_{k}^{(\mathrm{NL})} \rightarrow \frac{\tau_{c}-\tau}{\tau_{c}} \log _{2}\left(\frac{E_{u} M \beta \alpha_{k}}
 {  E_{u} \beta \alpha_{k}+ \frac{\sigma^{2}}{\tau}+\sigma^{2}\left(1+\frac{\sigma^{2}}{\tau E_{u} \beta \alpha_{k}}\right)}\right).
 \end{align}
\end{corollary}

Unlike Corollary \ref{corollary_single_user_MNN}, it is observed that the SNR in Corollary \ref{corollary_single_user_rayleigh} is an increasing function of $\tau$.
This observation again emphasizes that the LoS environment is favorable for RIS-aided single-user systems.

\section{Numerical Results}\label{section_6}
In the simulations, we consider that the distance between the user and the RIS is $d_{UI}=20$ m, and the distance between the RIS and the BS is $d_{IB}=700$ m. 
The distance between the user and the BS can be calculated from the geometric relationship as $\left(d_{k}^{\mathrm{UB}}\right)^{2}=\left(d_{\mathrm{IB}}-d_{\mathrm{UI}} \sin \left(\frac{\pi}{5} \right)\right)^{2}+\left(d_{\mathrm{UI}} \cos \left(\frac{\pi}{5} \right)\right)^{2}$. 
With the distances, large-scale factors are set to $ \alpha_{k}=10^{-3} d_{\mathrm{UI}}^{-2} $, $ \beta=10^{-3} d_{\mathrm{IB}}^{-2.5}  $ and $\gamma_{k}=10^{-3}\left(d_{k}^{\mathrm{UB}}\right)^{-4}$. Besides, we set $M=N=64$, $p=30$ dBm, $\sigma^2=-104$ dBm, $\tau_{c}=196$\cite{ngo2013energy}, $\tau=1$, $\delta=1$ and $\varepsilon_{k}=10$. 

\begin{figure}[htb]
	\setlength{\abovecaptionskip}{0pt}
	\setlength{\belowcaptionskip}{-20pt}
	\centering
	\includegraphics[width= 0.45\textwidth]{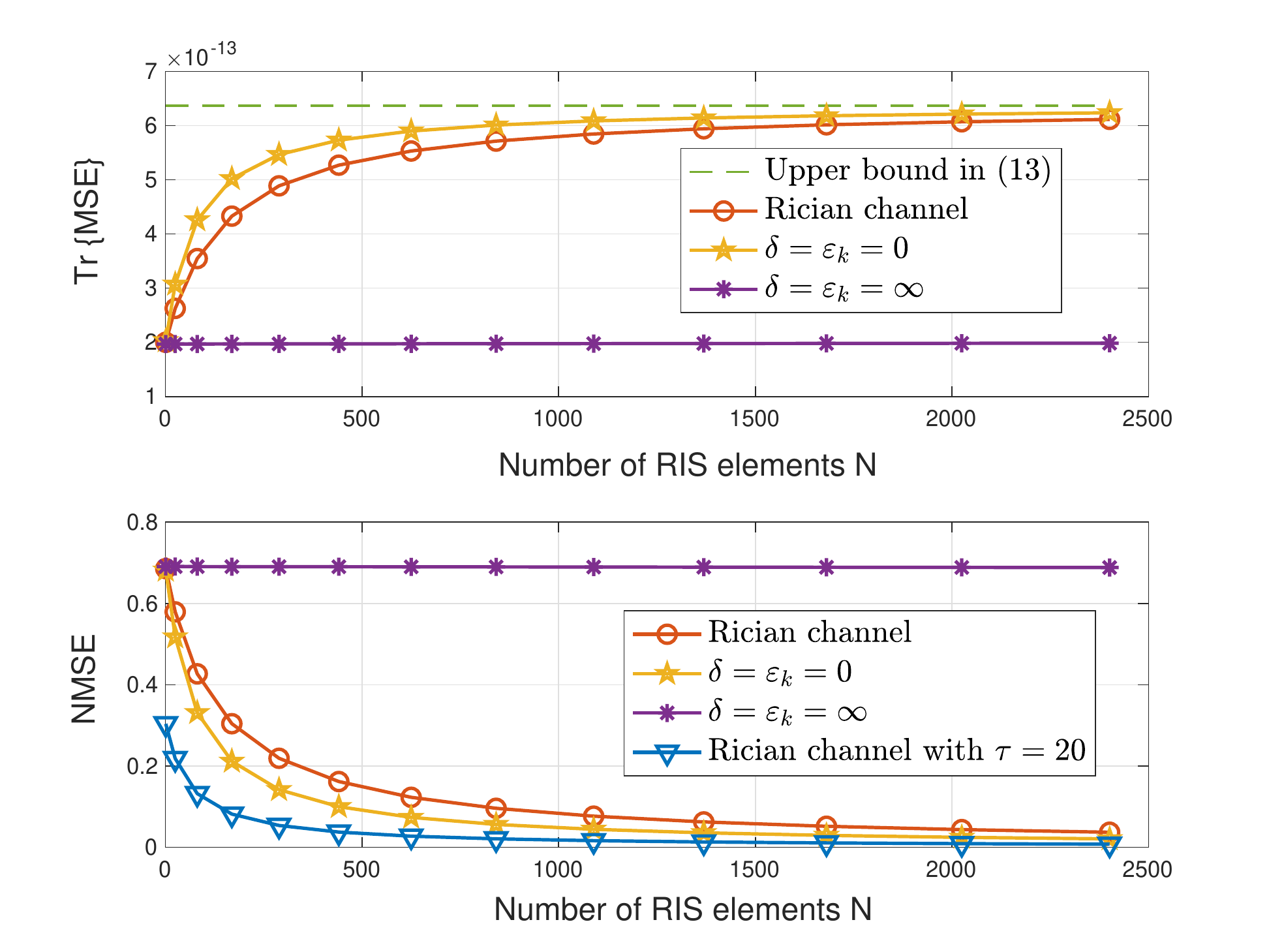}
	\DeclareGraphicsExtensions.
	\caption{MSE and NMSE versus the number of RIS elements.}
	\label{figure2}
\end{figure}
We first investigate the estimation quality of overall channel ${\bf Q}$. As our analysis, we can see that the MSE is an increasing function while the NMSE is a decreasing function with respect to $N$. With a large $N$,  the MSE approaches an upper bound and the NMSE converges to zero. In addition, the fully LoS case has the same performance with RIS-free traditional systems. Therefore, the MSE and NMSE in the fully LoS case do not change with $N$. Furthermore, compared with the NMSE with an increased $\tau$, we can see that increasing $N$ can incur a similar impact. 

\begin{figure}[htb]
	\setlength{\abovecaptionskip}{0pt}
	\setlength{\belowcaptionskip}{-20pt}
	\centering
	\includegraphics[width= 0.45\textwidth]{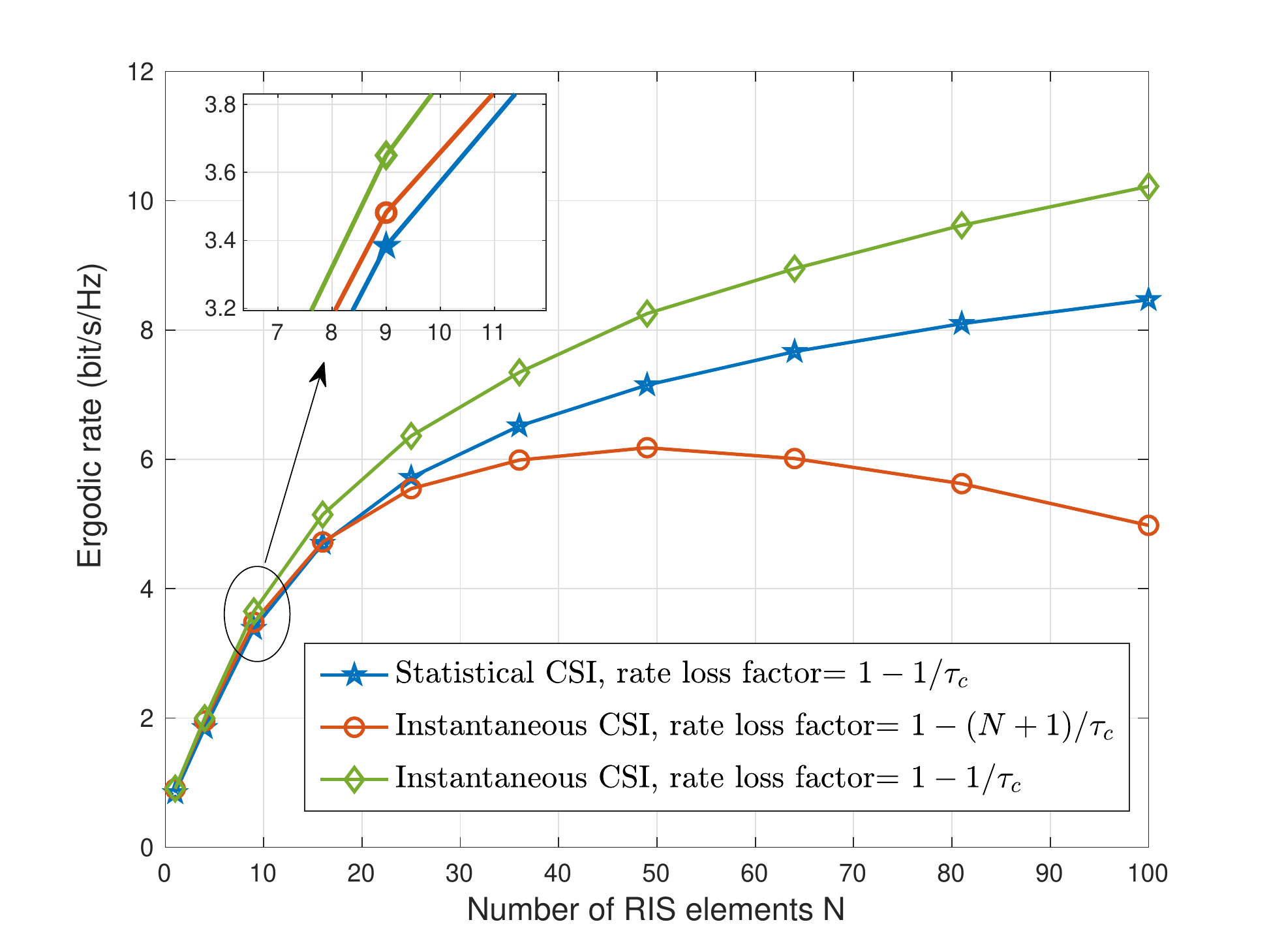}
	\DeclareGraphicsExtensions.
	\caption{Statistical CSI-based design versus instantaneous CSI-based design. }
	\label{figure2.5}
\end{figure}

In Fig. \ref{figure2.5}, we compare our design with instantaneous CSI-based design, which 
 estimates the instantaneous channels in each coherence interval as \cite[Section $ \rm\uppercase\expandafter{\romannumeral5} $]{9130088}. With single user, the minimum pilot length is $1+N$ in instantaneous CSI scheme\cite{9130088}. Therefore, its rate (red line) should be multiplied by a loss factor $ 1-\frac{1+N}{\tau_{c}}$. Meanwhile, we consider an ideal case (green line) with a loss factor $1-\frac{1}{\tau_{c}} $, which is the same loss factor of our scheme. We can see that if we ignore the additional rate loss in channel estimation (green line), using instantaneous CSI can achieve better performance, especially in large $N$ regime. However, when considering the overhead in pilot transmission (red line), the instantaneous CSI scheme first increase but then decrease, which performs worse than our design. This is because  instantaneous CSI scheme need longer time for channel estimation, which is proportional to $N$. Hence, when $N$ is large, only little time is left for data transmission.
However, our design can keep improving the rate performance as increasing $N$. Therefore, compared with instantaneous CSI-based design, statistical CSI-based design can realize better performance but with lower overhead and complexity in practical systems.

\begin{figure}[htb]
	\setlength{\abovecaptionskip}{0pt}
	\setlength{\belowcaptionskip}{-20pt}
	\centering
	\includegraphics[width= 0.43\textwidth]{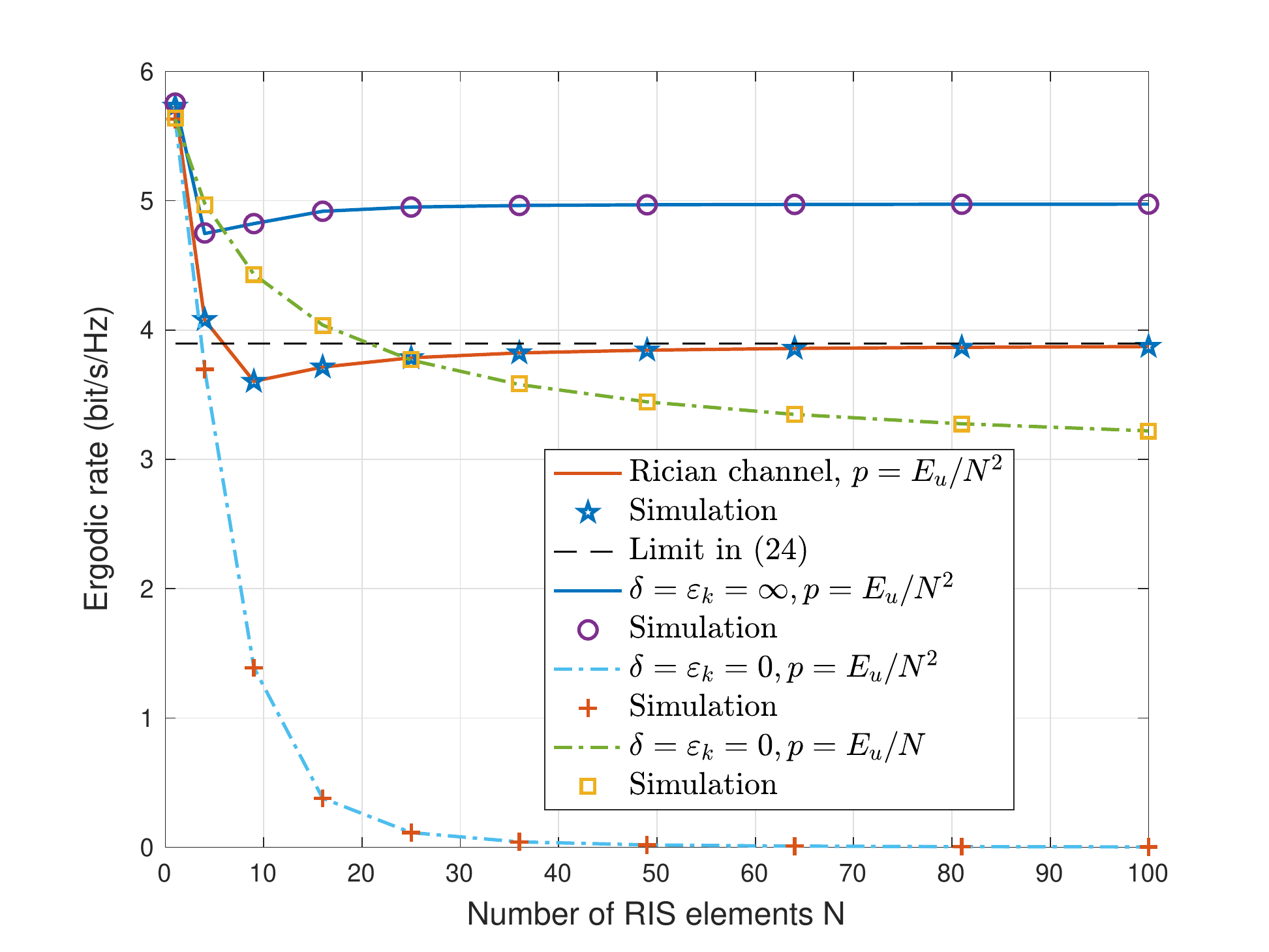}
	\DeclareGraphicsExtensions.
	\caption{ Rate versus $N$ with scaled-down transmit power $p=E_u/N^2$ or $p=E_u/N$, where $E_u=20$ dB.}
	\label{figure3}
\end{figure}

In Fig. \ref{figure3}, we characterizes the power scaling law with $N$. As presented in Corollary \ref{corollary_single_user_MNN}, the rate will converge to a limit if we reduce the power proportionally to $1/N^2$ under Rician channel, and this limit is maximized under fully LoS RIS-related channels. However, under NLoS RIS-related channels, reducing the power proportionally to $1/N^2$ will make the rate zero, and it can be reduced proportionally to $1/N$ as shown in Corollary \ref{corollary_single_user_rayleigh}. These observations validate that the LoS environment is preferable for RIS-aided single user systems.

\section{Conclusion}\label{section_7}
This paper has investigated RIS-aided MISO systems, taking the channel estimation error into consideration. An LMMSE estimator has been proposed to obtain the CSI for overall channels. Next, we have derived a closed-form expression for the ergodic date rate under the Rician channel model. Based on the derived expression, we have designed the phase shifts of the RIS based only on statistical CSI, which significantly reduces the signaling overhead and computational complexity of the system. We have also theoretically analyzed the power scaling laws under different kinds of channel conditions, which validates that the LoS environment is beneficial for RIS-aided MISO systems.

\bibliographystyle{IEEEtran}
\vspace{-6pt}
\bibliography{myref.bib}
\end{document}